\documentclass[conference,USletter]{IEEEtran}
\IEEEoverridecommandlockouts

\usepackage{amsmath,amssymb,amsfonts}
\usepackage{graphicx}
\usepackage{textcomp}
\usepackage{booktabs}
\usepackage{xcolor}
\usepackage[capitalize]{cleveref}
\usepackage[footnotesize]{caption}
\usepackage{subcaption}
\usepackage{url}
\usepackage[nolist]{acronym}

\def\BibTeX{{\rm B\kern-.05em{\sc i\kern-.025em b}\kern-.08em
    T\kern-.1667em\lower.7ex\hbox{E}\kern-.125emX}}

\def\mathlette#1#2{{\mathchoice{\mbox{#1$\displaystyle #2$}}%
                           {\mbox{#1$\textstyle #2$}}%
                           {\mbox{#1$\scriptstyle #2$}}%
                           {\mbox{#1$\scriptscriptstyle #2$}}}}
\renewcommand{\Vec}[1]{\mathlette{\boldmath}{#1}} 
 
\newcommand{\be}{\begin{equation}}
\newcommand{\ee}{\end{equation}}
\newcommand{\ba}{\begin{array}}
\newcommand{\ea}{\end{array}}
\newcommand{\bdm}{\begin{displaymath}}
\newcommand{\edm}{\end{displaymath}}
\newcommand{\bea}{\begin{eqnarray}}
\newcommand{\eea}{\end{eqnarray}}
\newcommand{\bean}{\begin{eqnarray*}}
\newcommand{\eean}{\end{eqnarray*}}

\graphicspath{ {Figures/} }
    
\begin{document}

\title{Ray Tracing Algorithm for Reconfigurable Intelligent Surfaces}

\author{ Sara Sandh, Hamed Radpour, Benjamin Rainer, Markus Hofer and Thomas Zemen\\
	AIT Austrian Institute of Technology, Vienna, Austria\\
	Email: thomas.zemen@ait.ac.at
 }

\maketitle

\begin{abstract}
Ray tracing accelerated with \acp{gpu} is an accurate and efficient simulation technique of wireless communication channels. In this paper, we extend a \ac{gpu}-accelerated \ac{rt} to support the effects of \acp{ris}. To evaluate the electric field, we derived a \ac{ris} path loss model that can be integrated into a \ac{rt} and enables further extensions for the implementation of additional features and incorporation into complex reflective scenarios. We verify the derivation and implementation of our model by comparison with empirical measurements in a lab environment. We demonstrate the capabilities of our model to support higher-order reflections from the \ac{ris} to the receiver. We find that such components have a significant effect on the received signal strength, concluding that the extensions of advanced functionality enabled by our model play an important role in the accurate modeling of radio wave propagation in an environment including \ac{ris}s.
\end{abstract}

\section{Introduction}
The propagation environment in wireless communication is traditionally viewed as an uncontrollable and unpredictable system that affects the quality of the signal as it interacts with the environment. Recently, however, the concept of \emph{smart radio environments}, that is, networks in which the environment actively assists in transferring and processing information, has emerged as a result of \acfp{ris}, i.e., an electronically controllable surface of \ac{em} material that can manipulate signals in real-time to constructively or destructively interfere at desired locations \cite{Basar19}.     

To achieve the full potential of \acp{ris}, several major challenges remain: real-time channel estimation, feedback of the received signal strength, control of the \ac{ris}, system optimization, and deployment, to name a few \cite{Basar19}. To understand the limitations and optimize configurations, it is thus of great importance to develop accurate \ac{ris} models and numerical radio wave simulation tools. 

In \cite{Tang22} a path loss model for \ac{ris} in the \ac{mmwave} frequency band is presented for an anechoic environment. Radpour et al. showed in \cite{Radpour23} a validation with empirical measurement data using a 37 element \ac{ris} in the mmWave band that has the ability to amplify the impinging signal at each element by $3$\,dB using an orthogonal polarization transform \cite{Wu22}.

For the practical use of \acp{ris} in indoor environments, the reflections of the radio signal on objects in the environment have to be taken into account. In \cite{Choi23} and \cite{xing2022raytracing}, the effect of \ac{ris}s are added to a \ac{rt} focusing on outdoor as well as indoor-to-outdoor scenarios. However, the higher-order reflection of the signal reflected by the \ac{ris} on other objects is not considered. In \cite{Huang22c}, \ac{rt} results for a \ac{ris} are shown for a frequency band below $6\,\text{GHz}$, without disclosing implementation details. The authors in \cite{hao2023extended} study the usage of a \ac{rt}-based \ac{ris} model in an urban scenario for a center frequency of 3.5 GHz for path loss-based system-level simulation. In \cite{Vitucci23}, the authors propose a macroscopic ray-based model to compute the reradiated wave from a \ac{ris}. However, the study of a \ac{ris}-assisted \ac{rt}-based model for indoor scenarios that include the reflected propagation paths between \ac{ris} and \ac{rx} is not yet available, to the best of the authors' knowledge.

\vspace{1em}
\noindent \emph{Scientific contributions of this paper:}
\begin{itemize}
    \item We derive a \ac{ris} propagation model based on the ideas of \cite{Tang22} to compute the electric field of a \ac{ris} in complex and challenging environments.
    \item We implement the model within a \ac{gpu}-accelerated \ac{rt} \cite{Rainer20}, based on the NVIDIA OptiX ray tracing engine. The implementation is verified against empirical measurement data from \cite{Radpour24} using a $127$ element \ac{ris}.
    \item We demonstrate how to consider the effect of the environment that is illuminated by the \ac{ris}, including reflected paths between the \ac{ris} and the \ac{rx}.
\end{itemize}

\section{Simulation of Reconfigurable Intelligent Surfaces}

The \ac{ris} considered in this paper is implemented using an array of sub-wavelength patch elements that receive and re-radiate the signal with a real-time adjustable reflection coefficient, thus applying a weighting and phase shift to the reflected wave. A simplified propagation analysis of a \ac{ris}-assisted communication channel can be reduced to the following steps \cite{Bjornson20}:

\begin{enumerate}
    \item Compute the emitted electric field from the \ac{tx} towards each \ac{ris} element.
    \item Determine the impinging field at each \ac{ris} element based on the free space path loss factor and antenna characteristics of the \ac{ris} element.
    \item Compute the reemitted field from each \ac{ris} element towards the \ac{rx} and other reflective objects in the environment using the image method.
    \item For each reemitted component, compute the electric field at the \ac{rx}.
    \item Apply the superposition principle to obtain the total electric field at the \ac{rx}.
\end{enumerate}

The \ac{gpu} accelerated RT considered in this paper implements a two-stage hybrid ray tracing method, combining analytic ray tracing and ray launching to model the channel characteristics as described in \cite{Rainer20}. In the first stage, potential reflection, diffraction, and diffuse interaction points are determined analytically, without extensive validation. In the second stage, we launch rays towards the potential interaction points using the NVIDIA OptiX ray tracing engine, ensuring that only non-blocked paths are considered at the \ac{rx}. 

We propose an approach for the integration of \ac{ris}s in our \ac{rt} by launching rays towards the defined patch antenna elements. Upon intersection with the \ac{ris}, the electric field is updated according to the path loss model derived in \cref{Implementation_ModelDerivation} and relaunched as described in \cref{Implementation_SupportingReflections}.

\subsection{Propagation Model}\label{Implementation_ModelDerivation}

Starting from the path loss model in \cite{Tang22} that is validated with measurements from an anechoic room in \cite{Radpour23}, we generalize the far-field electric field equations for better incorporation into the \ac{rt}. 

\begin{figure}[ht]
   \centering
   \includegraphics[width=\columnwidth]{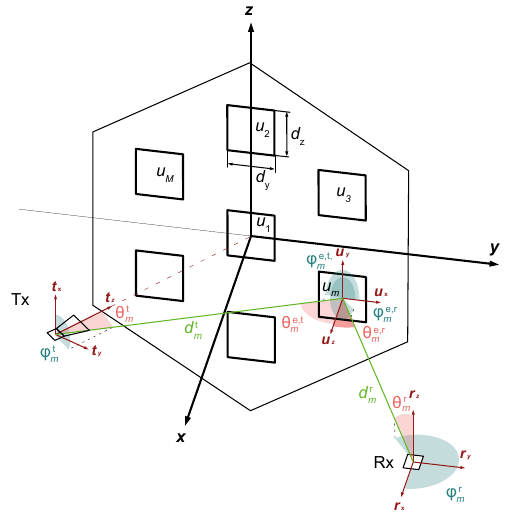}
   \caption{Illustration of a \ac{ris} with patch elements arranged in a hexagonal grid in the \textit{yz} plane. In the depicted scenario, a horn antenna at the transmitter radiates towards the center of the \ac{ris}, and a monopole \ac{rx} antenna captures the signal reflected from the \ac{ris}.}
   \label{fig:RISEnvironment}
\end{figure}

We define $d_y$ and $d_z$ as the \ac{ris} element effective dimensions, $G$ as the \ac{ris} element gain, and $\Gamma_m$ as the complex reflection coefficient of element $m$. The Euclidean distance from \ac{ris} element $m$ to the \ac{tx} is $d^\text{\,t}_m$ and $d^\text{\,r}_m$ to the \ac{rx}, as depicted in \cref{fig:RISEnvironment}.

The angles of the incoming and outgoing rays are given by ($\phi^\text{\,e,t}_m$, $\theta^\text{\,e,t}_m$) and ($\phi^\text{\,e,r}_m$, $\theta^\text{\,e,r}_m$), respectively, both in relation to the coordinate system of the \ac{ris} element. Throughout this paper, elevation $\theta$ is measured from the \textit{z} axis of the coordinate system in question, and azimuth $\phi$ from the \textit{x} axis. The angle of the emitted ray from the \ac{tx} and impinging wave on the \ac{rx} are given by ($\phi^\text{\,t}_m$, $\theta^\text{\,t}_m$) and ($\phi^\text{\,r}_m$, $\theta^\text{\,r}_m$), respectively, both measured from the corresponding antenna orientation matrix, as indicated in \cref{fig:RISEnvironment}.

The magnitude of the directional power density (Poynting vector) can be expressed in terms of the electric field as

\be
    S = \frac{|\Vec E|^2}{2\eta_0}\,, 
    \label{eq:PowerDensity_E}
\ee where $\eta_0=120 \pi \, \text{$\Omega$}$ is the impedance of free space. An electromagnetic wave emitted from an isotropic transmit antenna has the power density

\be
    S = \frac{P_\text{t}}{4\pi d^2} 
    \label{eq:PowerDensity_Pt}
\ee at a distance $d$ from the \ac{tx}, with $P_\text{t}$ denoting the transmit power. 

The magnitude of the electric field emitted from a \ac{tx} antenna with gain $G_\text{t}$ and normalized radiation pattern $F_\text{t}(\phi, \theta)$, can thus be expressed as
\be
    |\Vec E_\text{t}| = \sqrt{ \frac{2\eta_0 P_\text{t}}{4\pi d^2} G_\text{t} F_\text{t}(\phi,\theta)} 
    e^{\frac{-j2\pi d}{\lambda}} \,,
    \label{eq:Et_noniso}
\ee where azimuth $\phi_\text{t}$ and elevation $\theta_\text{t}$ give the direction of the emitted field measured from the \textit{z} axis of the \ac{tx} antenna orientation matrix, and $\Vec E_\text{t} \in \mathbb{C}^3$ is a three-dimensional complex vector describing the magnitude and polarization of the propagating electric field. 

In ray tracing, the distance $d$ is undefined until the ray hits an object or the \ac{rx}. Thus, the delay and attenuation of the signal is not accounted for when computing the initial electric field at the \ac{tx}. Rays launched towards the $m$-th \ac{ris} element center point thus have a field strength
\be
    |\Vec E^\text{\,t}_m| = \sqrt{ 2\eta_0 P_\text{t} G_\text{t} F_\text{t}(\phi^\text{\,t}_m, \theta^\text{\,t}_m)} \,,
    \label{eq:Et_nodist}
\ee
and the impinging electric field on element $m$ is given by
\begin{align}
    \Vec E^\text{\,RIS,in}_m & = 
    \sqrt{ 
        \frac{F(\phi^\text{\,e,t}_m, \theta^\text{\,e,t}_m) A_m}{4\pi (d^\text{\,t}_m)^2}
    } 
    e^{\frac{-j2\pi d^\text{\,t}_m}{\lambda}}
    \Vec{E}^\text{\,t}_m 
    \, .
    \label{eq:Erisin}
\end{align}

In addition to the delay and attenuation of the signal at distance $d^\text{\,t}_m$, \eqref{eq:Erisin} accounts for the element's normalized radiation pattern $F(\phi,\theta)$ in the direction of the impinging wave and the effective area $A_m=d_y d_z$ of the \ac{ris} element. The reemitted electric field from the $m$-th \ac{ris} element is given by 

\be
    \Vec E^\text{\,RIS,out}_m = \sqrt{G F(\phi^\text{\,e,r}_m, \theta^\text{\,e,r}_m) } \Gamma_m 
    \Vec  E^\text{\,RIS,in}_m \,,
    \label{eq:Erisout}
\ee accounting for the element's gain, normalized radiation pattern, and complex reflection coefficient. Equation \eqref{eq:Erisout} can additionally be extended to account for the \ac{ris}s effect on polarization. 

By combining \eqref{eq:Erisin} and \eqref{eq:Erisout}, we obtain an expression for each \ac{ris} element's impact on the electric field:

\begin{equation}
    \Vec E^\text{\,RIS}_m =  
    \sqrt{\frac{G
    F_\text{\,RIS}
    d_y d_z}{4\pi}}
    \frac{\Gamma_m}{d^\text{\,t}_m}
    e^{\frac{-j2\pi d^\text{\,t}_m}{\lambda}}
     \Vec{E}_{\text{t},m} \,,
    \label{eq:Eris}
\end{equation} 
where $F_\text{\,RIS}:=F(\phi^\text{\,e,t}_m, \theta^\text{\,e,t}_m)F(\phi^\text{\,e,r}_m, \theta^\text{\,e,r}_m)$.

Accounting for the free space path loss, \ac{rx} normalized radiation pattern $F_\text{r}(\phi,\theta)$, and effective area $A_\text{r}=\frac{\lambda^2}{4\pi}G_\text{r}$, the electric field at the \ac{rx} is given by
\begin{align}
\begin{split}
    \Vec E^\text{\,r}_m & =  
    \sqrt{
        \frac{A_\text{r} F_\text{r}(\phi^\text{\,r}_m,\theta^\text{\,r}_m)}
        {4\pi (d^\text{\,r}_m)^2}
    } e^{\frac{-j2\pi d^\text{\,r}_m}{\lambda}}
    \Vec E^\text{\,RIS}_m
    \\
     &= \sqrt{ G_\text{r} F_\text{r}(\phi^\text{\,r}_m,\theta^\text{\,r}_m) }
    \frac{\lambda}{4\pi d^\text{\,r}_m}
    e^{\frac{-j2\pi d^\text{\,r}_m}{\lambda}}
    \Vec E^\text{\,RIS}_m \,.
    \label{eq:Er}
\end{split}
\end{align}

The superposition of the received components from each element gives the total electric field at the \ac{rx}
\begin{equation}
    \Vec E_\text{r} = \sum^{M}_{m=1} \Vec E^\text{\,r}_m 
    \label{eq:Etot}
\end{equation}
and the received signal power
\begin{equation}
    P_\text{r} = \frac{|\Vec E_\text{r}|^2}{2\eta_0} \,.
    \label{eq:Pr}
\end{equation}

When modeling a \ac{ris} in an anechoic chamber with blocked \ac{los} between the \ac{tx} and \ac{rx}, \eqref{eq:Et_nodist} and \eqref{eq:Eris} to \eqref{eq:Pr} reduce to the path loss model proposed by Tang et al. in \cite{Tang22}, assuming that there is no polarization mismatch, that element gain $G=\frac{4\pi d_y d_z}{\lambda^2}$, and that antennas and \ac{ris} are configured as described in \cite{Tang22}. In our generalized model, equations \eqref{eq:Et_nodist}, \eqref{eq:Eris}, and \eqref{eq:Er} provide a complete description of the propagation steps for a \ac{ris} in a wireless channel, and can be combined with other propagation models.

\subsection{Supporting Reflections}\label{Implementation_SupportingReflections}
In a reflective environment, the radiation pattern of the \ac{ris} illuminates other objects and may reflect towards the \ac{rx} and has a significant impact on the received signal power and \ac{cir}. As each \ac{ris} element reemits the signal and thus acts as a \ac{tx}, we can employ the image method as described in \cite{Gan15} and \cite{Quatresooz21} to identify reflection points between the \ac{ris} and the \ac{rx}. Similarly, the image method can be used to identify reflected components from the \ac{tx} impinging on the \ac{ris}, or reflections from the \ac{ris} impinging on itself.

\begin{figure}[ht]
    \centering
    \includegraphics[width=\columnwidth]{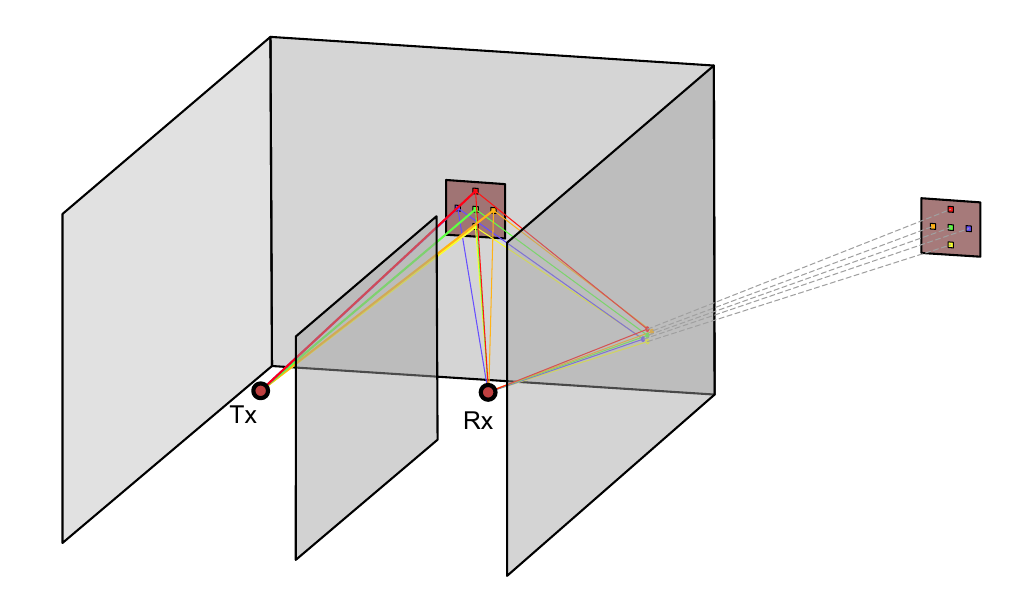}
    \caption{Procedure for identifying valid reflection paths between the \ac{ris} and \ac{rx} using the image method.}
    \label{fig:ExtendRT_Reflections}
\end{figure}

To efficiently identify valid reflection paths between the \ac{ris} and the \ac{rx}, we search for valid permutations of surfaces forming unblocked paths, only between the \ac{ris} center point and the \ac{rx} position. Thereafter, we perform the image method on each \ac{ris} element only for the valid and unblocked surface permutations, as seen in \cref{fig:ExtendRT_Reflections}. The electric field of reemitted rays from the \ac{ris} are computed according to \eqref{eq:Eris}, and according to the Fresnel equations upon specular reflection \cite{Collin92}.

\section{Simulation Setup}
\label{sec:SimSetup}

In this paper, we focus on an industrial automation scenario in which a directional \ac{tx} is positioned in such a way that it maximizes the signal emitted towards the \ac{ris}, assuming that the direct path between the \ac{tx} and \ac{rx} may be blocked. The \ac{ris} is configured to maximize the outgoing signal for a mobile monopole \ac{rx} antenna moving in the \textit{xy} plane, for instance, mounted on a robot arm. The outgoing angle of the \ac{ris} is thus adapted to the position of the \ac{rx} in the \textit{xy} plane. We consider a reflective environment in which metallic walls were mounted on the sides and opposite of the \ac{ris} to demonstrate the potential effects of a reflective industrial environment, for instance, caused by surrounding machines and other large, metallic objects. 

To verify the derivation and implementation of our model, we performed a simulation equivalent to the measurement setup described in \cite{Radpour24}, conducted in a lab environment in which reflections are partially suppressed. The modeled scenario considers an active \ac{ris} with $M=127$ patch antenna elements operating in an anechoic chamber, thus only considering direct paths from the \ac{tx} to the \ac{rx} through the \ac{ris}.

In a realistic environment, however, the impact of \acp{mpc} can significantly affect the received signal power and \ac{cir}, despite the high attenuation from free space path loss and interaction with objects for \ac{mmwave} signals. Signal components from an active \ac{ris} on reflective objects will have a substantial contribution to the behavior of the channel.

To evaluate the impact of reflected components from the \ac{ris}, we simulate an additional scenario that models a reflective environment, and considers every \ac{mpc} from the \ac{ris}, including up to second-order reflections between the \ac{ris} and the receiver. The \ac{ris} is configured to maximize the signal towards an \ac{rx} at $\Vec r=(1.4\text{\,m},10^\circ,106^\circ)$, emitted from a \ac{tx} at $\Vec t=(1.86\text{\,m},-36^\circ,90^\circ)$, using spherical coordinates $(r,\phi,\theta)$ in relation to the \ac{ris}, as depicted in \cref{fig:RISEnvironment}. The simulation parameters are summarized in \cref{tab:SimulateReflective_Parameters}.
\begin{table}
\begin{center}
\caption{Configuration parameters and simulation setup for \ac{ris} in a reflective environment. }
\label{tab:SimulateReflective_Parameters}
\begin{tabular}{ll} 
\toprule
\textbf{Parameter}	      	               &  \textbf{Definition}\\
\midrule 
$f = 23.8$\,GHz		                       & 	 center frequency \\
\midrule
$P_\text{t}=10$\,dBm		               & 	 \ac{tx} transmit power\\
$G_\text{t} = 19$\,dB           &  \ac{tx} antenna gain\\
$G_\text{r} = 0$\,dB           &  \ac{rx} antenna gain\\
$\Vec t = (1.86\text{\,m},-36^\circ,90^\circ)$  & \ac{tx} position in relation to \ac{ris} \\
$\Vec r = (1.4\text{\,m},10^\circ,106^\circ)$ & \ac{rx} position in relation to \ac{ris} \\
\midrule
$M=127$                                     &    number of \ac{ris} elements\\
$d_y$, $d_z = 6.6$\,mm	                    & 	 effective \ac{ris} element size \\
$\Gamma_m \in \{1.25,\angle 0^\circ),(0,\angle 0^\circ)\}$   & reflection coefficients \\
\midrule
$\varepsilon_{r,\text{metal}}=1.0$ & relative permittivity of metal \\
$\sigma_\text{metal}=10^7 \frac{\text{S}}{\text{m}}$ & conductivity of metal \\
\bottomrule
\end{tabular}
\end{center}
\end{table}

\subsection{Reflective Environment}
To model the described environment in the \ac{rt}, three metallic walls and a \ac{ris} object are defined, as shown in \cref{fig:ResultsReflection_Rays}. Blocks of metallic material construct a $2\times 2.4\times 1$\,m room with an opening in which a $12\times12$\,cm \ac{ris} is placed. The \ac{ris} is centered at $\Vec{u}=(0,0,0.5)\,$m with surface normal $\Vec{n}=(1,0,0)$. The metallic material is defined as a \ac{pec} with relative permittivity $\varepsilon_r=1.0$ and conductivity $\sigma=10^7 \frac{\text{S}}{\text{m}}$. To omit direct components from the \ac{tx} to the \ac{rx}, a $0.9\times 0.05\times 1$\,m blocking panel is defined, centered at $(1.55,-0.475,0.5)\,$m in the geometry of the scene.

\subsection{Antenna Implementation}
The measurements \cite{Radpour24} consider a \ac{tx} horn antenna and a monopole \ac{rx} antenna. The \ac{tx} is implemented with a directional rotation symmetric antenna pattern 
\be 
    F_\text{t}(\theta)=(\cos{\theta})^{\frac{G_\text{t}}{2}-1}\,,
\ee with antenna gain $G_\text{t}=19$\,dB, and transmit power $P_\text{t}=10$\,dBm. The antenna orientation is set such that the main lobe at $\theta=0$ is focused towards the \ac{ris} center point $\Vec{u}$. The \ac{rx} antenna is implemented with a monopole antenna model 
\be 
    F_\text{r}(\theta)=\Bigg(\frac{
    \cos{\frac{2\pi h}{\lambda}\cos{\theta}}-\cos{\frac{2\pi h}{\lambda}}
    }
    {\sin{\theta}}\Bigg)^2\,,
\ee with antenna gain $G_\text{r}=0\,$dB and monopole height $h=\frac{\lambda}{4}$.

The orientation and position of the \ac{tx} remain constant throughout the simulation, whereas the position of the \ac{rx} is updated to sample the received signal at every position in the ranges $0.92\leq x\leq 1.52$\,m and $0.02\leq y\leq 0.92$\,m at $z=11.4$\,cm, as marked in \cref{fig:ResultsReflection_Rays}, with a sampling distance of $1$\,cm.

\subsection{Technical Details of the RIS}
The active \ac{ris}s considered in the reflective environment consists of $M=127$ patch antenna elements with an effective element size $d_y=d_z=6.6$\,mm, arranged in a hexagonal grid. We model the normalized radiation pattern of each element as 
\be
F(\theta)=\cos \theta\,, 
\ee in accordance with \cite{Tang22}. Each element is either switched on or off, providing an amplification of $1.25$ if switched on. The reflection coefficient alphabet is thus
\be
    \Gamma_m \in \{(1.25,\angle 0^\circ),(0,\angle 0^\circ)\} \, ,
    \label{eq:SimulateReflective_ActiveAlphabet}
\ee and elements are configured to optimize the received signal strength at the intended \ac{rx} position according to \cite[Algorithm 1]{Radpour23}.

\section{Results}
\begin{figure}
    \centering
    \includegraphics[width=\columnwidth]{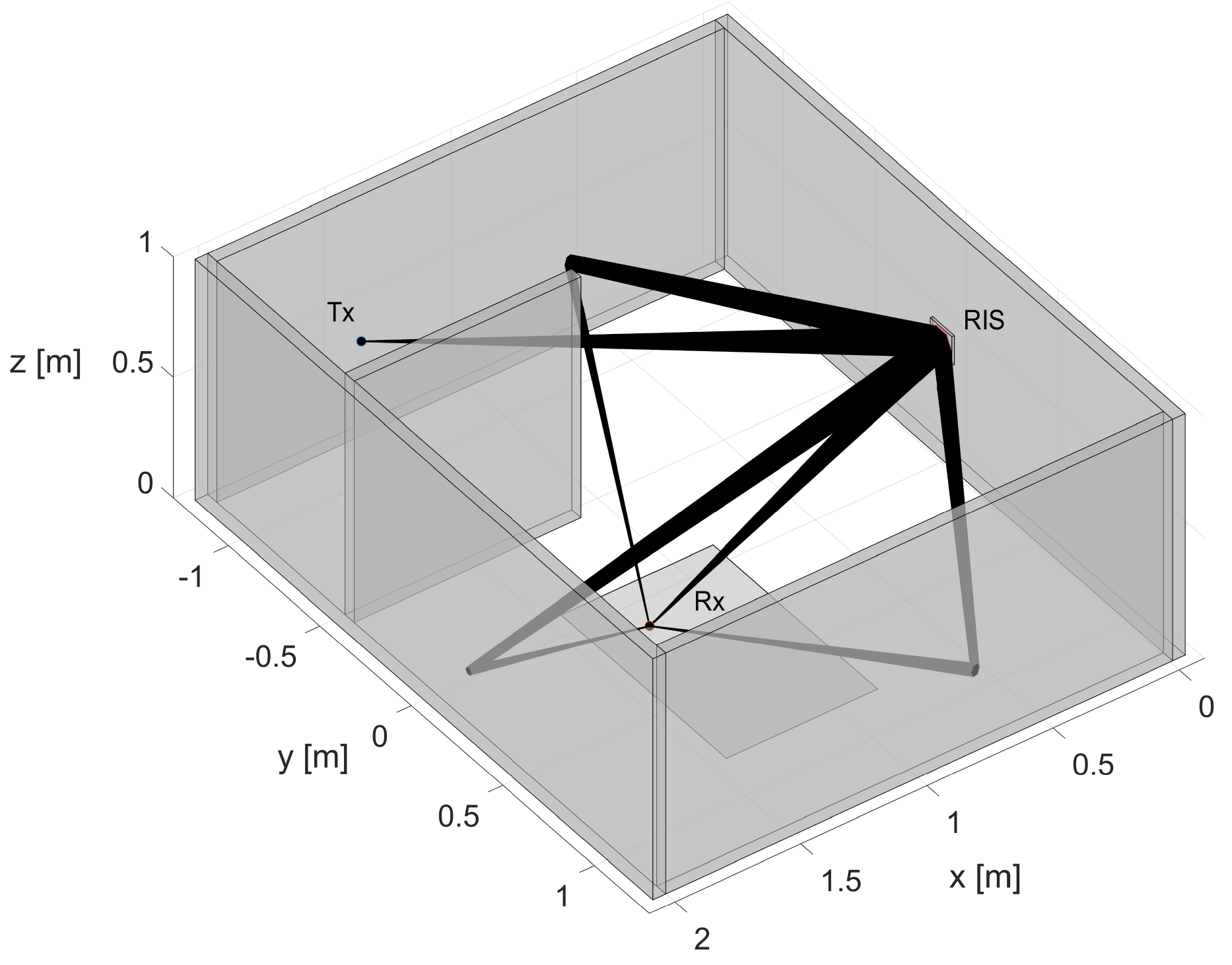}
    
    \caption{Path visualization for a \ac{ris} with $127$ elements in a reflection environment with the \ac{tx} at $\Vec t = (1.5,-1.1,0.5)$\,m, \ac{rx} at $\Vec r = (1.33,0.23,0.11)$\,m, and \ac{ris} centered at $\Vec{u}=(0,0,0.5)$\,m.}
    \label{fig:ResultsReflection_Rays}
\end{figure}
In this section, we validate the implemented model against measurements from \cite{Radpour24}. Furthermore, we present the obtained results of the \ac{rt} in a reflective environment by evaluating the obtained \acp{mpc} and received signal power in a simulation where reflections up to the second order from the \ac{ris} are taken into account.
\begin{figure*}[ht]
    \centering
    \begin{subfigure}[b]{0.66\columnwidth}
         \centering
         \includegraphics[width=\columnwidth]{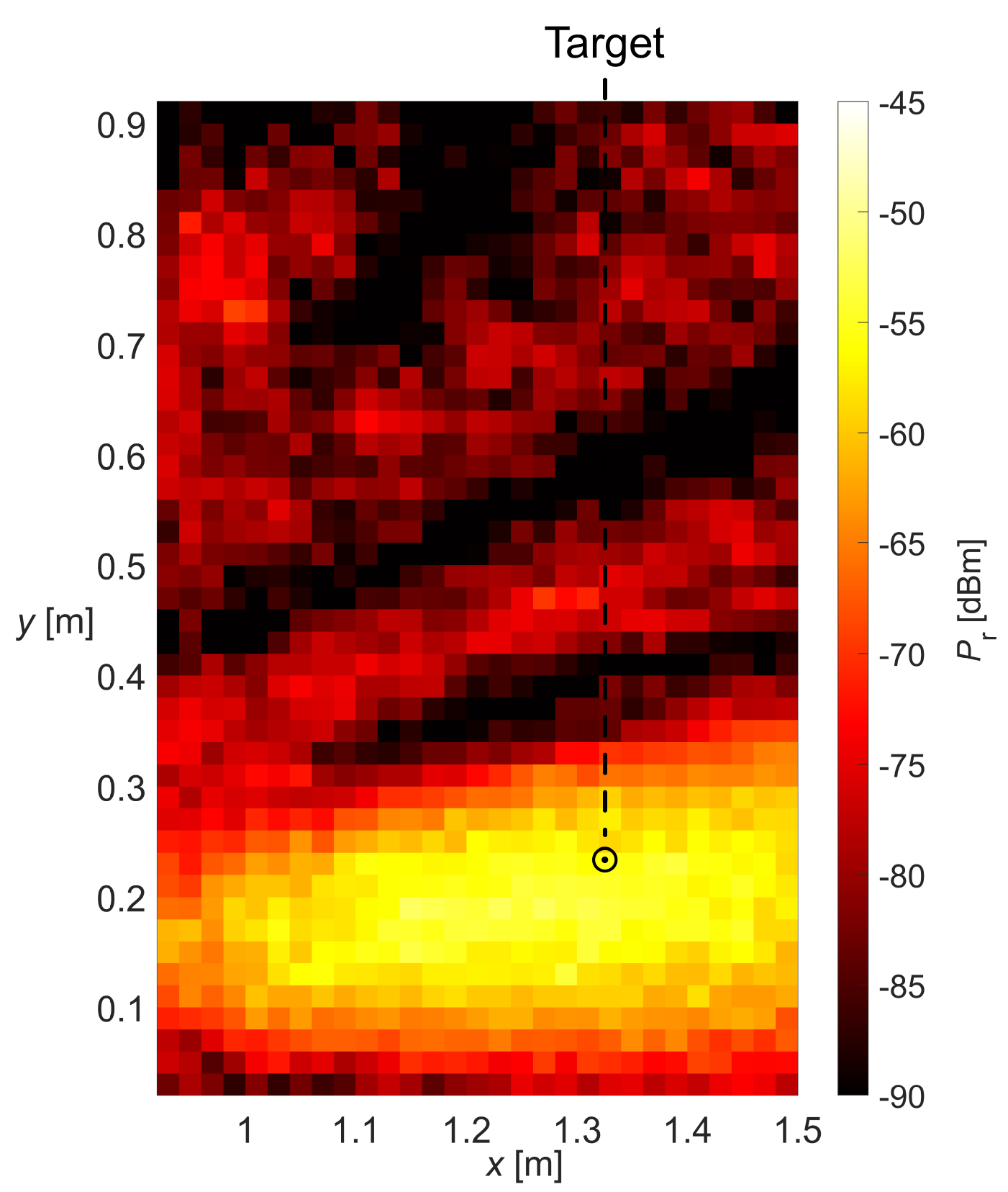}
         \caption{Measurements from \cite{Radpour24}.}
         \label{fig:Results_Measured}
    \end{subfigure}
    \begin{subfigure}[b]{0.66\columnwidth}
         \centering
         \includegraphics[width=\columnwidth]{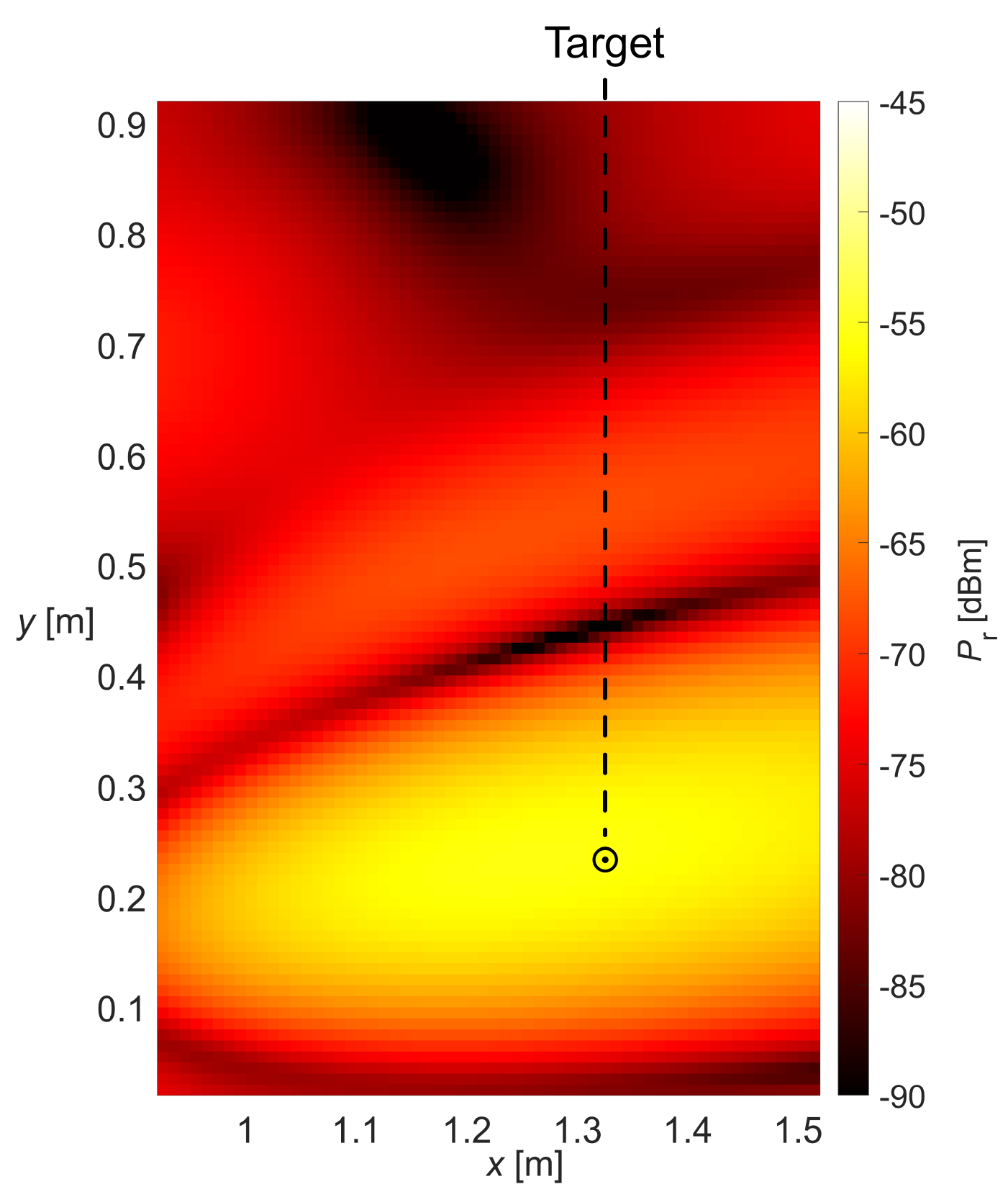}
         \caption{Simulation in anechoic room.}
         \label{fig:Results_NoReflections}
    \end{subfigure}
    \begin{subfigure}[b]{0.66\columnwidth}
         \centering
         \includegraphics[width=\columnwidth]{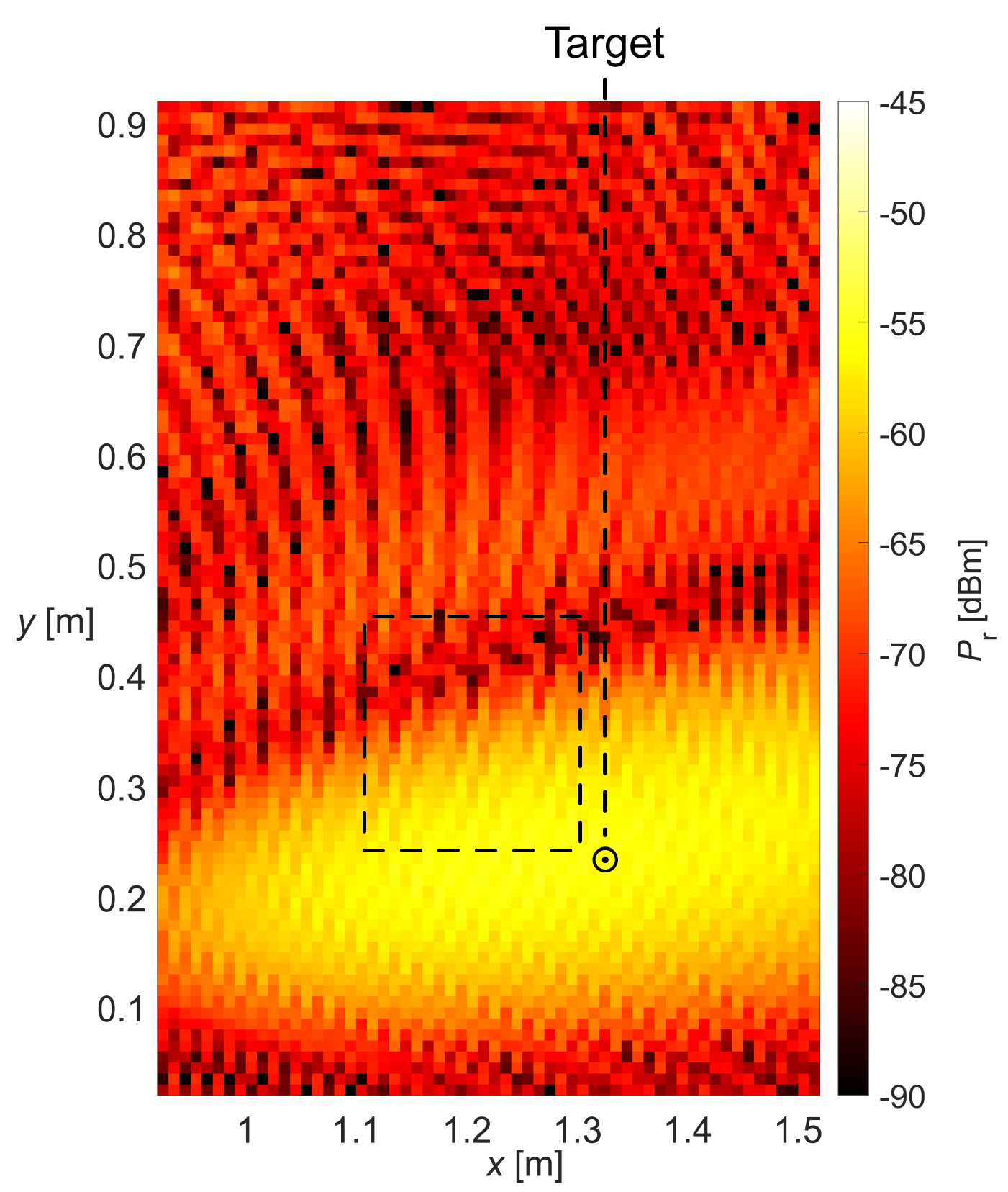}
         \caption{Simulation with higher-order reflections.}
         \label{fig:Results_Reflections}
    \end{subfigure}
    \caption{Received signal power $P_\text{r}$ from an active \ac{ris} over a region in the \textit{xy} plane, measured in a lab environment from \cite{Radpour24} in (a), simulated in an anechoic chamber (b), and simulated in a reflective environment (c). The \ac{ris} is configured to maximize the received signal at $\Vec r=(1.33,0.23,0.11)$. }
    \label{fig:Results_Validation}
\end{figure*}

\subsection{Ray Launching}
To evaluate the reflected paths between the \ac{ris} and \ac{rx}, we inspect the ray launching using a path visualization tool. The simulation which did not consider higher-order reflections launch exactly one ray towards each of the $127$ \ac{ris} elements. Upon intersection with the \ac{ris}, the rays are relaunched towards the \ac{rx}. The simulation considering higher-order reflections additionally identify reflection points towards which rays are relaunched after intersection with the \ac{ris}, which are thereafter specularly reflected until arrival at the \ac{rx}. \cref{fig:ResultsReflection_Rays} shows the generated paths from the simulation considering up to the second order reflections from the \ac{ris} for an \ac{rx} placed at $\Vec r=(1.33,0.23,0.11)\,$m.

\subsection{Received Signal Power}
The measurement from \cite{Radpour24} and equivalent \ac{rt} simulations yield the received signal powers plotted in \cref{fig:Results_Validation}. The received signal power is computed and plotted for every sampling point in the defined range by applying \eqref{eq:Etot} and \eqref{eq:Pr} to every \ac{mpc} that arrives at the \ac{rx}.

In \cref{fig:Results_Measured,fig:Results_NoReflections} the signal power is plotted for the measurement in a lab environment with suppressed reflections and a simulation in an anechoic chamber, respectively. The \ac{ris} placed at $\Vec u = (0,0,0.5)$ is configured to reflect the signal towards the \ac{rx} at $\Vec r=(1.33,0.23,0.11)$. The measured scenario shows a dominant beam of approximately $0.5\times0.15$\,m, with a received signal power from $-60$ to $-55$\,dBm, and weaker side lobes from $-80$ to $-75$\,dBm. For a detailed description of the measurement results, we refer to \cite{Radpour24}. Simulation with the \ac{rt} achieves similar results, with a dominant beam centered around the \ac{rx} with signal power from approximately $-60$ to $-55$\,dBm over a $0.5\times0.2\,$m region, and weaker side lobes $-80$ to $-70$\,dBm. 

The simulation considering reflections from the \ac{ris} up to the second order generate the plot in \cref{fig:Results_Reflections}. The constructive and destructive interference from reflected components is evident, and comparison with \cref{fig:Results_NoReflections} shows that reflected propagation paths from the \ac{ris} to the \ac{rx} have a large impact on the received signal strength at several locations in the considered range.

\begin{figure}[t!]
    \centering
    \includegraphics[width=\columnwidth]{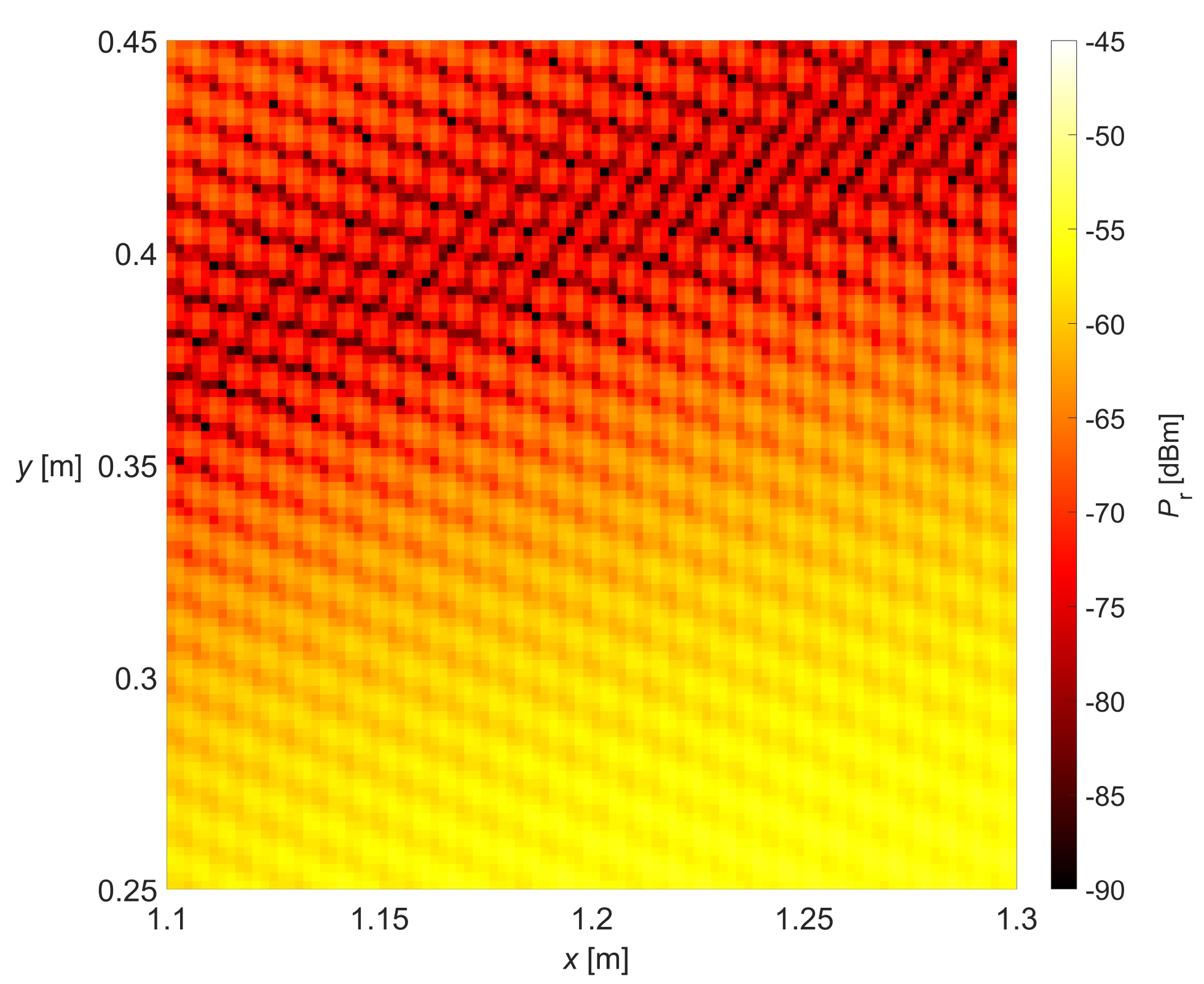}
    \caption{High-resolution plot of received signal power $P_\text{r}$ from a \ac{rt} simulation of an active \ac{ris} in a reflective environment, considering direct and up to second order reflected signal components from the \ac{ris}.}
    \label{fig:Results_Closeup}
\end{figure}

To visualize the small-scale fading pattern in more detail, we performed a high-resolution simulation of a smaller section of the \textit{xy} plane, between $1.1\leq x\leq 1.3$\,m and $0.25\leq y\leq 0.45$\,m, as marked in \cref{fig:Results_Reflections}, with a sampling distance of $2\,$mm. The received signal power is plotted in \cref{fig:Results_Closeup}, clearly showing the small-scale fading with constructive and destructive interference at approximately every $\frac{\lambda}{2}$. The reflections from the \ac{ris} can have a large impact on the received signal power, showing the importance of considering them for an accurate simulation of \acp{ris}, particularly in reflective environments.

Accurate channel descriptions are essential for the development of \ac{ris} control algorithms. Our \ac{rt} can simulate realistic and complex environments in which reflected components from the \ac{ris} are considered, enabling the development of \ac{ris} control algorithms accounting for the small-scale fading. 

\section{Conclusion}
By manipulating the signal to constructively or destructively interfere at desired locations, \acp{ris} offer promising solutions to the propagation issues for \acp{mmwave}, making it an important enabling technology for 6G wireless networks. The development of \acp{ris} require accurate models and numerical simulation tools, ray tracing being a popular technique for efficient wireless channel modeling. 

In this paper, we proposed an approach for supporting the effects of \acp{ris} in complex environments by employing the image method to compute reflected paths between the \ac{ris} and the \ac{rx}. For the evaluation of the electric field, we adapted the \ac{ris} path loss model proposed by Tang et al. in \cite{Tang22} for a \ac{gpu}-accelerated ray tracer. 

We validated our implementation against measurements in a lab environment from \cite{Radpour24}. The correctness of our implementation was verified, and the model yielded reliable results for the simulated and measured case. We further evaluated the effects of a \ac{ris} in a reflective environment, showing that our model successfully enabled the extension to advanced propagation mechanisms. Furthermore, we found that higher-order reflected components from the \ac{ris} on the environment have a significant effect on the received signal strength, indicating that the extensions enabled by our model play an important role in the accurate modeling of \acp{ris}. 

\section*{Acknowledgement}
The work of S. Sandh, H. Radpour, B. Rainer, M. Hofer and T. Zemen (corresponding author) has been funded within the Principal Scientist grant Dependable Wireless 6G Communication Systems (DEDICATE 6G) at the AIT Austrian Institute of Technology.


\begin{thebibliography}{10}

\bibitem{Basar19}
E.~Basar, M.~Di~Renzo, J.~De~Rosny, M.~Debbah, M.-S. Alouini, and R.~Zhang,
  ``{Wireless Communications Through Reconfigurable Intelligent Surfaces},''
  {\em IEEE Access}, vol.~7, pp.~116753--116773, 2019.

\bibitem{Tang22}
W.~Tang, X.~Chen, M.~Z. Chen, J.~Y. Dai, Y.~Han, M.~D. Renzo, S.~Jin, Q.~Cheng,
  and T.~J. Cui, ``{Path Loss Modeling and Measurements for Reconfigurable
  Intelligent Surfaces in the Millimeter-Wave Frequency Band},'' {\em IEEE
  Transactions on Communications}, vol.~70, no.~9, pp.~6259--6276, 2022.

\bibitem{Radpour23}
H.~Radpour, M.~Hofer, L.~W. Mayer, A.~Hofmann, M.~Schiefer, and T.~Zemen,
  ``{Active Reconfigurable Intelligent Surface for the Millimeter-Wave
  Frequency Band: Design and Measurement Results},'' in {\em IEEE Wireless
  Communications and Networking Conference (WCNC), Dubai, United Arab
  Emirates}, April 2024, to be presented, online available
  {https://arxiv.org/abs/2306.04515}.

\bibitem{Wu22}
L.~Wu, K.~Lou, J.~Ke, J.~Liang, Z.~Luo, J.~Y. Dai, Q.~Cheng, and T.~J. Cui,
  ``{A Wideband Amplifying Reconfigurable Intelligent Surface},'' {\em IEEE
  Transactions on Antennas and Propagation}, vol.~70, no.~11, pp.~10623--10631,
  2022.

\bibitem{Choi23}
H.~Choi, J.~Oh, J.~Chung, G.~C. Alexandropoulos, and J.~Choi, ``{WiThRay: A
  Versatile Ray-Tracing Simulator for Smart Wireless Environments},'' {\em IEEE
  Access}, 2023.

\bibitem{xing2022raytracing}
Y.~Xing, F.~Vook, E.~Visotsky, M.~Cudak, and A.~Ghosh, ``{Raytracing-Based
  System Performance of Intelligent Reflecting Surfaces at 28 GHz},'' in {\em
  ICC 2022-IEEE International Conference on Communications}, pp.~498--503,
  IEEE, 2022.

\bibitem{Huang22c}
J.~Huang, C.-X. Wang, Y.~Sun, J.~Huang, and F.-C. Zheng, ``{A Novel Ray Tracing
  Based 6G RIS Wireless Channel Model and RIS Deployment Studies in Indoor
  Scenarios},'' in {\em 2022 IEEE 33rd Annual International Symposium on
  Personal, Indoor and Mobile Radio Communications (PIMRC)}, pp.~884--889,
  IEEE, 2022.

\bibitem{hao2023extended}
L.~Hao, S.~Schwarz, and M.~Rupp, ``{The Extended Vienna System-Level Simulator
  for Reconfigurable Intelligent Surfaces},'' in {\em 2023 Joint European
  Conference on Networks and Communications \& 6G Summit (EuCNC/6G Summit)},
  IEEE, 2023.

\bibitem{Vitucci23}
E.~M. Vitucci, M.~Albani, S.~Kodra, M.~Barbiroli, and V.~Degli-Esposti, ``An
  efficient ray-based modeling approach for scattering from reconfigurable
  intelligent surfaces,'' {\em IEEE Transactions on Antennas and Propagation},
  2024.

\bibitem{Rainer20}
B.~Rainer, D.~L{\"o}schenbrand, S.~Zelenbaba, M.~Hofer, and T.~Zemen,
  ``{Towards a Non-Stationary Correlated Fading Process for Diffuse Scattering
  in Ray Tracing},'' in {\em IEEE 31st Annual International Symposium on
  Personal, Indoor and Mobile Radio Communications}, 2020.

\bibitem{Radpour24}
H.~Radpour, M.~Hofer, D.~L{\"o}schenbrand, L.~W. Mayer, A.~Hofmann,
  M.~Schiefer, and T.~Zemen, ``{Reconfigurable Intelligent Surface for Indoor
  Industrial Automation: mmWave Propagation Measurement, Simulation, and
  Control Algorithm Requirements},'' in {\em IEEE International Symposium on
  Personal, Indoor and Mobile Radio Communications (PIMRC), Valencia, Spain},
  September 2024, submitted, online available
  {https://arxiv.org/abs/2402.04844}.

\bibitem{Bjornson20}
E.~Bj{\"o}rnson, {\"O}.~{\"O}zdogan, and E.~G. Larsson, ``{Reconfigurable
  Intelligent Surfaces: Three Myths and Two Critical Questions},'' {\em IEEE
  Communications Magazine}, vol.~58, no.~12, pp.~90--96, 2020.

\bibitem{Gan15}
M.~Gan, {\em {Accurate and Low-complexity Ray Tracing Channel Modeling}}.
\newblock PhD thesis, Technical University of Vienna, 2015.

\bibitem{Quatresooz21}
F.~Quatresooz, S.~Demey, and C.~Oestges, ``{Tracking of Interaction Points for
  Improved Dynamic Ray Tracing},'' {\em IEEE Transactions on Vehicular
  Technology}, vol.~70, no.~7, pp.~6291--6301, 2021.

\bibitem{Collin92}
R.~Collin, {\em {Foundations for Microwave Engineering}}.
\newblock IEEE Press Series on Electromagnetic Wave Theory, McGraw-Hill, 1992.

\end{thebibliography}

\begin{acronym}[RIS]
\acro{5g}[5G]{fifth-generation cellular network}
\acro{iot}[IoT]{internet of things}
\acro{m2m}[M2M]{machine-to-machine}
\acro{6g}[6G]{sixth-generation cellular network}
\acro{mmwave}[mmWave]{millimeter-wave frequency}
\acro{ris}[RIS]{reconfigurable intelligent surface}
\acro{em}[EM]{electromagnetic}
\acro{snr}[SNR]{signal-to-noise ratio}
\acro{nlos}[NLOS]{non-line-of-sight}
\acro{fem}[FEM]{finite element method}
\acro{fdtd}[FDTD]{finite-difference time-domain}
\acro{gpu}[GPU]{graphics processing unit}
\acro{ait}[AIT]{AIT Austrian Institute of Technology}
\acro{tx}[TX]{transmitter}
\acro{rx}[RX]{receiver}
\acro{los}[LOS]{line-of-sight}
\acro{io}[IO]{interacting object}
\acro{mpc}[MPC]{multi-path component}
\acro{rt}[RT]{ray tracer}
\acro{utd}[UTD]{uniform theory of diffraction}
\acro{sbr}[SBR]{shooting and bouncing ray}
\acro{cpu}[CPU]{central processing unit}
\acro{zag}[ZAG]{z-aligned axis geometry}
\acro{osm}[OSM]{OpenStreetMap}
\acro{ctf}[CTF]{channel transfer function}
\acro{cir}[CIR]{channel impulse response}
\acro{pdp}[PDP]{power delay profile}
\acro{dsd}[DSD]{Doppler spectral density}
\acro{pec}[PEC]{perfect electrical conductor}
\acro{rmse}[RMSE]{root-mean-square error}
\acro{mape}[MAPE]{mean absolute percentage error}
\acro{mse}[MSE]{mean square error}
\acro{mmwave}[mmWave]{milimeter wave}
\end{acronym}

\end{document}